\documentclass[runningheads,orivec]{llncs}

\usepackage[T1]{fontenc}
\usepackage{microtype}
\usepackage{graphicx}
\usepackage{hyperref}
\usepackage{url}
\usepackage{listings}
\usepackage{xcolor}
\usepackage{tikz}
\usetikzlibrary{arrows.meta, positioning, shapes.geometric, calc}

\begin{document}

\title{Reconstructing OPC UA Address Spaces\\from Time-Series Databases\thanks{This is the authors' accepted version of a paper accepted at AI4IP~2026 (workshop at DEXA~2026). The final authenticated version will appear in the Springer CCIS series; the version of record will be available via its DOI once published.}}
\titlerunning{Reconstructing OPC UA Address Spaces from Time-Series Databases}

\author{Lukas Lürzer\orcidID{0009-0000-5953-1381} \and
        Hannes Unger\orcidID{0000-0003-3715-6845} \and
        Stefan Huber\orcidID{0000-0002-8871-5814}}
\authorrunning{L. Lürzer et al.}
\institute{
    Josef Ressel Centre for Intelligent and Secure Industrial Automation,\\
    Salzburg University of Applied Sciences, Austria\\
    \email{\{lukas.luerzer,hannes.unger,stefan.huber\}@fh-salzburg.ac.at}
}

\maketitle

\begin{abstract}
OPC~UA has become the dominant open protocol in operational technology. Time-series
databases routinely archive OPC~UA telemetry but discard the semantic metadata
(node hierarchy, engineering units, and type definitions) which gives sensor
values their meaning. Recovering this information from a time-series
database is non-trivial: namespace indices recorded at the source are session-local
and unstable across restarts, and naive merging across multiple source servers
results in identifier collisions. We present \emph{opcua-ts}, an
implemented architecture that persists this semantic information alongside
its telemetry in a general-purpose time-series database under a
lifecycle-stable join key, and that reconstructs the source address space
as a live OPC~UA endpoint. We characterize the conditions under which the
reconstruction is sound across multi-source deployments and validate the
approach with a NodeSet2 XML round-trip against the source server. Initial
results from a boiler-simulator round-trip indicate that the approach is
feasible.
\end{abstract}

\section{Introduction}

OPC~UA~\cite{opcua-spec} has become the dominant protocol for machine-to-machine
communication in operational technology (OT). Every compliant server exposes an
\emph{address space}: a typed, hierarchical graph of objects and variables with
engineering units, data types, references, and browse paths. Additionally, the OPC Foundation maintains a growing family of standardized
\emph{companion specifications} (for robotics, machine tools, process control, energy,
and other domains), so that compliant servers across vendors expose semantically
interoperable type systems out of the box.

In practice this information model is usually lost at the OT/IT boundary,
because typical ingestion pipelines (Telegraf, MQTT bridges, ad-hoc
adapters) marshal each value into a schema-light row keyed by node
identifier, and the surrounding structural metadata has no natural home
there. Downstream analytics, dashboards, and ML pipelines therefore operate
on context-free numeric streams and reintroduce the missing semantics
through hand-maintained mapping tables, which drift out of sync as the
source address space evolves and the mapping is updated only when someone
notices.

Recovering the address space from such a time-series store is non-trivial.
The binding between values and their semantic context is broken at ingestion,
namespace indices recorded in the source server are session-local and unstable across server
restarts, and naive merging across multiple source servers admits identifier
collisions in both the namespace and the browse-name dimension. We therefore
investigate the following research questions:

\begin{itemize}
  \item \textbf{RQ1.} Can the OPC~UA information model be persisted alongside its
  telemetry in a general-purpose time-series database with a join key?
  \item \textbf{RQ2.} Under what conditions can the original address space be
  reconstructed from such a store and served as a live OPC~UA endpoint, including
  across multi-source deployments where namespace and browse-name collisions are
  unavoidable?
  \item \textbf{RQ3.} How can the semantic fidelity of a reconstructed address space
  be validated against the original source?
\end{itemize}

To answer these questions, we contribute (i)~a join key derived from the
namespace \emph{URI} rather than the session-local namespace index, which
makes it canonical (one form per node) and lifecycle-stable (unchanged
across server restarts and crawl boundaries), and under which telemetry
and semantic metadata can be co-located in a general-purpose time-series
store; (ii)~an aggregate reconstruction server that builds the persisted
information model into a live OPC~UA endpoint and manages namespace and
browse-name collisions across multiple sources; and (iii)~a validation
procedure based on a NodeSet2~\cite{opcua-spec} round-trip against the
source server.

\section{Related Work}

Historical access to OPC~UA data is standardized through the Historical
Access profiles defined in OPC~UA Part~11~\cite{opcua-part11}, and a number
of commercial historians implement them. These products treat the address
space as a runtime structure that is mirrored at configuration time, not as
a first-class artifact persisted alongside the time series; an address space
recovered from such a historian therefore reflects the configuration
snapshot at mirror time rather than the lifecycle of the source server.

Semantic-aware digital-twin platforms address the closely related problem of
giving industrial values a queryable semantic context, but typically do so
through proprietary graph or semantic stores attached to the time-series
layer via integrations, rather than by co-locating the semantic and
temporal layers under a shared key. To our knowledge, no open system
persists OPC~UA semantic metadata in a general-purpose time-series database
such that the source address space can be reconstructed programmatically.

\section{System Architecture}

\begin{figure}[t]
\centering
\resizebox{\columnwidth}{!}{%
\begin{tikzpicture}[
  font=\scriptsize,
  >={Latex[length=1.5mm]},
  src/.style   ={draw, rounded corners, minimum width=20mm, minimum height=7mm,
                 align=center, fill=gray!10,   inner sep=1pt},
  comp/.style  ={draw, rounded corners, minimum width=18mm, minimum height=8mm,
                 align=center, fill=blue!8,    inner sep=1pt},
  store/.style ={draw, rounded corners, minimum width=22mm, minimum height=10mm,
                 align=center, fill=orange!15, inner sep=1pt},
  cli/.style   ={draw, rounded corners, minimum width=16mm, minimum height=8mm,
                 align=center, fill=green!12,  inner sep=1pt},
  lbl/.style   ={font=\tiny, midway, above, inner sep=1pt}
]
  \node[src] (s1) {OPC~UA server\\(machine A)};
  \node[src, below=2mm of s1] (s2) {OPC~UA server\\(machine B)};
  \node[src, below=2mm of s2] (s3) {\dots};

  \node[comp, right=10mm of s1] (crawl)    {Semantic\\crawler};
  \node[comp, right=10mm of s3] (telegraf) {Telegraf};

  \node[store, right=10mm of crawl, yshift=-9mm] (db)
        {Time-series store\\\textit{values + model}};

  \node[comp, right=10mm of db]    (recon) {Reconstruction\\server};
  \node[cli,  right=10mm of recon] (c1)    {OPC~UA\\clients};

  \coordinate (fanC) at ($(crawl.west)+(-4mm,0)$);
  \coordinate (fanT) at ($(telegraf.west)+(-4mm,0)$);
  \draw[->] (s1.east) -- (fanC) -- (crawl.west);
  \draw[->] (s2.east) -| (fanC) -- (crawl.west);
  \draw[->] (s3.east) -| (fanC) -- (crawl.west);
  \draw[->] (s1.east) -| (fanT) -- (telegraf.west);
  \draw[->] (s2.east) -| (fanT) -- (telegraf.west);
  \draw[->] (s3.east) -- (fanT) -- (telegraf.west);

  \draw[->] (crawl.east)    -| node[lbl,pos=0.25] {model}  (db.north);
  \draw[->] (telegraf.east) -| node[lbl,pos=0.25] {values} (db.south);

  \draw[->] (db)    -- node[lbl] {reload} (recon);
  \draw[->] (recon) -- node[lbl] {OPC~UA} (c1);
\end{tikzpicture}%
}
\caption{Architecture proposal of \emph{opcua-ts}: a semantic crawler and Telegraf persist the
information model and live values under a shared join key; a reconstruction server
serves the aggregate as a live OPC~UA endpoint.}
\label{fig:goal}
\end{figure}
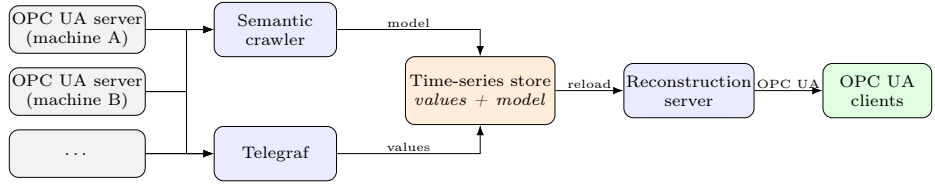

Figure~\ref{fig:goal} shows the workflow. Two ingestion paths feed the
store under a shared \texttt{node\_id} tag: a \emph{semantic crawler}
persists per-node metadata, while \emph{Telegraf}~\cite{telegraf-opcua}
polls live values. The design is source-agnostic: any compliant OPC~UA
server (PLC-embedded, gateway, simulator) can be used, and the
reconstruction server reproduces its information model without requiring
access to the original device.

\subsection{Storage Model}

\emph{opcua-ts} co-locates two complementary stores in the same
time-series database, joined by a shared key: a \emph{values store}
carrying live telemetry, written by Telegraf, and a \emph{metadata store}
carrying the semantic information model, written by the crawler. The two
stores share a canonical join key \texttt{node\_id} (its grammar appears
in Listing~\ref{lst:nodeid}). The key is derived from the namespace
\emph{URI} rather than the namespace \emph{index}: indices are
session-local and unstable across server restarts, whereas URIs are
durable across the lifecycle of the source server.

The metadata store carries, per node, the display name, browse path,
data-type name, type-definition name, parent \texttt{node\_id}, engineering
unit, \texttt{EURange}, non-hierarchical references, and a monotonic
\texttt{crawl\_version} field that enables both \texttt{last()} snapshots
and historical change detection without overwriting prior rows.
Listing~\ref{lst:nodeid} shows the join-key grammar and a representative
metadata record.

\begin{lstlisting}[caption={Canonical \texttt{node\_id} and a sample \texttt{opcua\_metadata}
row (selected fields).}, label={lst:nodeid}, float=t]
# Canonical node_id grammar
<namespace_uri>|<id_type>|<identifier>
# Example:  http://myplant.com/boiler/UA/|i|1002

# Sample opcua_metadata row (selected fields):
node_id       : http://myplant.com/boiler/UA/|i|1002
display_name  : Loop Temperature
browse_path   : /Objects/Boiler1/BoilerLoop1/LoopTemperature
data_type     : Double
type_def      : AnalogItemType
eu_range      : -50:200
engineering_unit : degree Celsius
parent_node_id   : http://myplant.com/boiler/UA/|i|1001
crawl_version    : 42
\end{lstlisting}

\subsection{Crawler Pipeline}

The crawler performs a depth-first traversal of the source server starting
from the standard \texttt{Objects} (\texttt{i=85}) and \texttt{Types}
(\texttt{i=86}) roots, recording for each node the hierarchical references
used to reach it and the \texttt{HasTypeDefinition} target. A second pass
enriches \texttt{Variable} nodes with \texttt{EURange} and
\texttt{EngineeringUnits} via the \texttt{HasProperty} reference.

\subsection{Reconstruction Pipeline}

The reconstruction server loads the latest metadata rows from the
metadata store into a node-opcua~\cite{node-opcua} address space. It
employs two collision-resolution mechanisms for multi-source deployments:
when two source servers share a namespace URI, the aggregate registers
each instance with a \texttt{\#source=} fragment suffix so that namespace
indices remain distinct; and a router places one container object per
source endpoint under \texttt{Objects}, isolating browse-name collisions.
Each \texttt{Variable} is wired to a value cache keyed on
\texttt{endpoint::node\_id}, populated by a background poll against the
values store, so clients see the most recent value with its original
\texttt{SourceTimestamp}.

\section{Validation via NodeSet2 Round-Trip}

We say a reconstructed address space is \emph{semantically faithful} to its
source if every browse, read, and type-resolution observation a standard
OPC~UA client may perform yields the same answer against the reconstruction
as against the source, and if the per-namespace NodeSet2 XML serialization
preserves \texttt{BrowseName}, \texttt{NodeClass}, \texttt{DataType},
\texttt{TypeDefinition}, browse-path position, engineering metadata, and
non-hierarchical references node by node. We assess this by exporting both
the source and the reconstructed server's address spaces to NodeSet2 XML
and comparing them attribute by attribute.

NodeSet2 XML is the canonical OPC~UA address-space serialization defined in
the OPC~UA specifications~\cite{opcua-spec}, and the node-opcua library
provides both a loader (\texttt{generateAddressSpace}) and a per-namespace
serializer (\texttt{toNodeset2XML()}). We rely on this serializer for both
directions of the round-trip.

\paragraph{Procedure.}
The round-trip proceeds in four steps. First, we export the source server's
namespace to NodeSet2 XML directly via \texttt{toNodeset2XML()}. Second, we
run the crawler against the same source server, populating the metadata
store with one row per node. Third, we instantiate a reconstruction from
the latest crawl. Finally, we export the reconstruction's namespace using
the same serializer and compare the two NodeSet2 files attribute by
attribute.

\paragraph{Initial findings.}
We exercised the round-trip on a boiler-simulator address space with eleven
functional nodes, including five \texttt{AnalogItemType} variables. The
source serialized to 25.4\,KB and the reconstruction to 19.9\,KB. All
functional nodes survive the round-trip together with their data types,
type definitions, the \texttt{EURange}, \texttt{EngineeringUnits}, and
\texttt{InstrumentRange} properties of the analog items, and the
non-hierarchical \texttt{GeneratesEvent} reference. This indicates that
the approach is feasible in principle.

\paragraph{Known gaps.}
The 5.5\,KB size delta between the two files is fully accounted
for by two optional display hints, \texttt{Definition} and
\texttt{ValuePrecision}. The metadata loader filters all
\texttt{PropertyType} sub-nodes from the snapshot, because node-opcua's
\texttt{addAnalogDataItem()} re-emits the \texttt{EURange},
\texttt{EngineeringUnits}, and \texttt{InstrumentRange} properties from
the primary metadata fields, and re-loading the source's own property
nodes would collide. The two hints are dropped as collateral; widening the
metadata query to preserve them is straightforward future work.

\section{Discussion and Conclusion}

Coupling the semantic and temporal layers in a single store lets analytics
pipelines join telemetry with engineering units and browse paths without an
external mapping table, allows the reconstruction server to expose a
semantically-rich OPC~UA endpoint to clients that cannot query the
time-series store directly, and makes the appearance or disappearance of a
\texttt{node\_id} (via \texttt{crawl\_version}) a queryable event.
Limitations remain: the reconstruction is snapshot-based, and custom
\texttt{ObjectType} / \texttt{VariableType} definitions reattach to
built-in base types in the current implementation. OPC~UA semantic
metadata can therefore be persisted alongside its telemetry without
sacrificing the ability to reconstruct the source address space. Future
work covers type-hierarchy preservation and subscription-driven crawls.

\begin{credits}
\subsubsection{\ackname}
During the preparation of this work, the authors used AI tools for language
editing and formatting assistance.
\end{credits}

\bibliographystyle{splncs04}
\bibliography{references}

\end{document}